\documentclass{ws-procs9x6}

\def\etal {{\it et al.}}

\begin{document}

\title{An odd (parity) test of Lorentz symmetry with atomic dysprosium}

\author{Nathan Leefer$^*$ and Michael Hohensee}

\address{Physics Department, University of California,\\
Berkeley, CA 94720-7300, USA\\
$^*$E-mail: naleefer@berkeley.edu}

\begin{abstract}
We propose using a Stark interference technique to directly measure the odd-parity $c_{0j}$ components of the electron sector $c_{\mu\nu}$ tensor of the Standard-Model Extension.  This technique has been shown to be a sensitive probe of parity violation in atomic dysprosium in a low-energy, tabletop experiment, and may also be straightforwardly applied to test Lorentz invariance. We estimate that such an experiment may be sensitive to $c_{0j}$ coefficients as small as $10^{-18}$.
\end{abstract}

\bodymatter

\section{Introduction}

Tests of Lorentz invariance in atomic systems typically hinge on detecting small energy shifts of bound states of elementary particles, due to the coupling of a Lorentz violating field to either the particle spin~\cite{Brown2010,Gemmel2010} or momentum.~\cite{Smiciklas2011,Hohensee2013} Spectroscopic measurement of the energy difference between the nearly degenerate, opposite-parity states [Xe]$4f^{10}5d6s$, $J=10$ (state A) and [Xe]$4f^{9}5d^{2}6s$, $J=10$ (state B) in atomic dysprosium (Dy) has been shown to be a sensitive probe of the parity-even $c_{jk}$ components of the Standard-Model Extension's (SME) electron-sector $c_{\mu\nu}$ tensor~\cite{Hohensee2013}.  These coefficients shift the energy of bound electrons  in direct proportion to the expectation value of $\hat{p}_{j}\hat{p}_{k}$, where $\hat{p}_{j}$ is the sum of the $j$-components of their momentum.  This energy shift modulates as a function of the electrons' orientation and velocity with respect to an inertial frame, and their position in an external gravitational potential~\cite{Lane1999,KosteleckyTasson2010,HohenseeMuellerTBP}.  Such experiments are best suited as probes of the rotational symmetry-breaking $c_{JK}$ components of the $c_{\mu\nu}$ tensor in the Sun-centered celestial equatorial frame (SCCEF), but are also indirectly sensitive to the parity-odd $c_{TJ}$ terms, which contribute to the terrestrial laboratory-frame observable $c_{jk}$ in proportion to the Earth's boost velocity $\beta\approx 10^{-4}$ with respect to the Sun~\cite{Kostelecky2002}.  Currently, the most stringent bounds on the parity-odd $c_{TJ}$ components are kinematic constraints from the observation of highly boosted laboratory~\cite{Altschul2010} or astrophysical~\cite{Altschul2006v1and2} sources.

We propose a low-energy atomic physics experiment that would be directly sensitive to the parity-odd components of the $c_{\mu\nu}$ tensor. The nonrelativistic quantum Hamiltonian derived in Ref.~\refcite{Kostelecky1999} includes the term
\begin{equation}
\delta h = \left(-a_j + m e_j + m c_{(0j)}\right) \frac{p_j}{m}c,
\label{sec2:eq1}
\end{equation}
where $m$ is the electron mass, $c$ is the speed of light, $c_{(0j)}$ indicates the symmetric combination $c_{0j}+c_{j0}$, and $p_{j}=\sum_{n=1}^{N}p_{j}^{(n)}$ is the total electron momentum projection along the $j$-axis.  The $a_{j}-me_{j}\equiv (a_{\rm eff})_{j}$ component of this term is unobservable in non-gravitational tests of flavor-conserving systems, as it can be removed from the fully relativistic theory by redefinition of the electron wave function's global phase~\cite{Colladay1997and1998}.  Note however that should these terms describe a field with a non-metric coupling to gravity, they may be observable in experiments performed within a gravitational potential~\cite{KosteleckyTasson2010}.  This possibility will be explored elsewhere.  Here, we will focus on the $c_{(0j)}$ coefficients, which give rise to observable effects in non-gravitational experiments.  These terms are typically ignored in spectroscopic tests of Lorentz invariance because the expectation value of the odd-parity operator $\hat{p}_{j}$ is zero for any bound state.  These terms do, however, contribute to the Hamiltonian matrix element between two states of opposite parity, and weakly drive transitions between different states.  The $c_{(0j)}$ terms can be thus be measured using experimental methods developed to detect small parity-violating matrix elements induced by electron-nucleon interactions mediated by $Z_{0}$ bosons~\cite{Khriplovich1991,Nguyen1997}.
\begin{figure}[t]
\begin{center}
\psfig{file=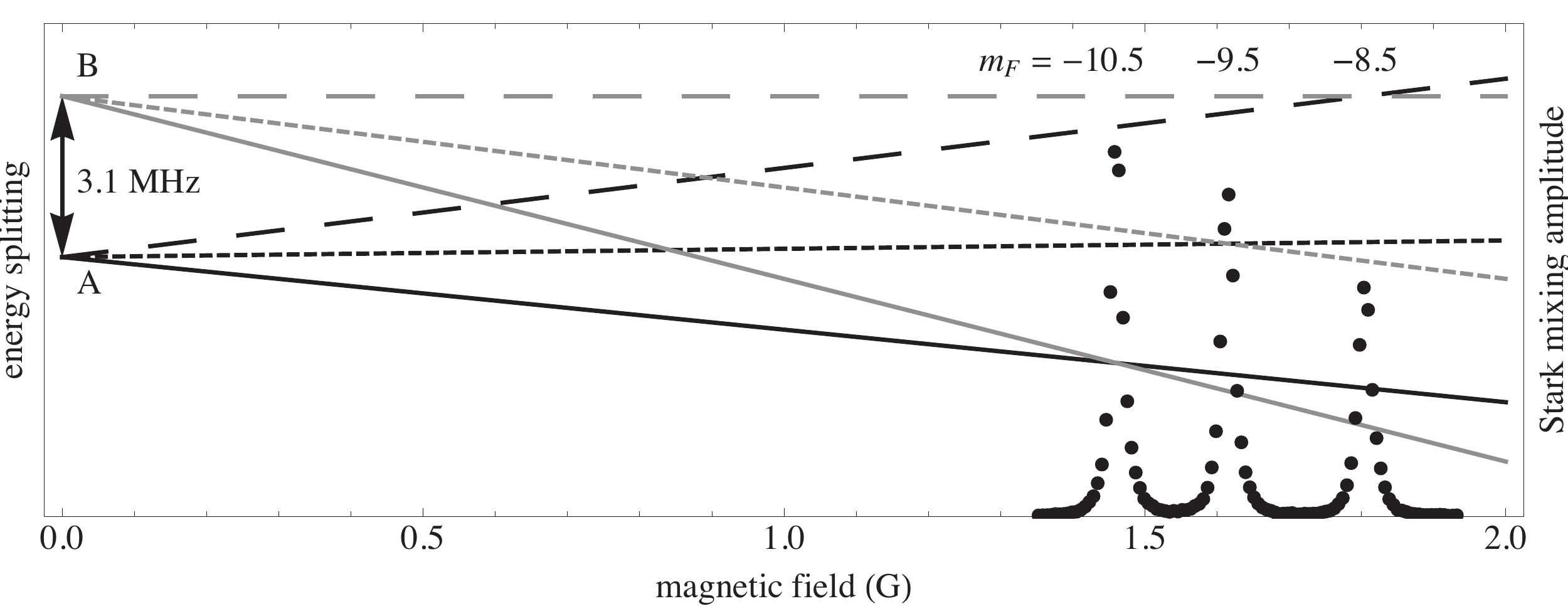,width=4.5in}
\end{center}
\caption{An example Zeeman crossing spectrum for Dy atoms in the presence of an $\omega/(2\pi) = 10$~kHz, $E_0 = 2.5$~V/cm electric field. Solid and dashed lines show the Zeeman shifts of three magnetic sublevels of states $A$ and $B$, relative to the $m_F = -8.5$ sublevel of state $B$, in $^{163}$Dy. The filled circles show resonances in the Stark induced mixing amplitudes when sublevels of equal $m_F$ are near degenerate.}
\label{sec2:fig1}
\end{figure}

The Stark interference technique used to measure parity violation in atoms relies on observing interference between the transition amplitude induced by parity-violation and that induced by an oscillating electric field~\cite{Noecker1988,Nguyen1997}. Consider again the opposite-parity states in dysprosium. In $^{163}$Dy the nuclear spin is $I = 5/2$ and the energy difference between the $F = 10.5$ hyperfine levels of state $A$ and state $B$ is only 3.1~MHz. The difference of Land\'{e} g-factors between states $A$ and $B$ allows sublevels of equal angular-momentum projection, $m_F$, to be brought to complete degeneracy with a modest magnetic field of less than 2~G, as shown in Fig.~\ref{sec2:fig1}.  Near these degeneracies, the Hamiltonian for the two near-degenerate states, coupled to one another by an applied electric field $E(t)=E_{0}\cos{\omega t}$, reduces to that of a two level system, and can be represented by the matrix
\begin{equation}
H = \begin{pmatrix} -i \gamma_A/2 &  E(t)d + i (H_{w}-c_{(0j)}W_{j}) \\  E(t)d -i (H_{w}-c_{(0j)}W_{j}) & \Delta \end{pmatrix},
\label{sec2:eq2}
\end{equation}
where $\gamma_A$ is the inverse lifetime of state $A$ ($B$ is assumed to be metastable), $\Delta$ is the residual energy difference between the states, $d$ is the electric-dipole matrix element connecting the states, $H_{w}$ is the conventional parity-violating matrix element between the two states, $W_{j} = \langle A,m_{F} |c\partial_{x_{j}}|B,m_{F}\rangle$, and sums are implied on repeated indexes.  This Hamiltonian is identical to that of Ref.~\refcite{Nguyen1997}, where the Lorentz-violating $c_{(0j)}W_{j}$ augments the parity-odd matrix element $H_w$.  Although the odd-parity matrix element $H_{w}+c_{(0j)}W_{j}$ is too small to significantly drive transitions by itself, it can interfere with the electric field induced transition amplitude, modifying the probability of the $B$ $\rightarrow$ $A$ transition with a term proportional to $E_{0}d(H_{w}-c_{(0j)}W_{j})$.  The detailed procedure for measuring $H_w$, and hence $c_{(0j)}W_{j}$, can be found in Ref.~\refcite{Nguyen1997} and is not reproduced here. The only distinction is that $c_{(0j)}W_{j}$ can be modulated by rotation of the atoms' quantization axis or of the laboratory frame, which makes it distinguishable from conventional parity violating signals.

To constrain $c_{(0j)}$ with this method requires calculation of the $W_{j}$ matrix elements, which in turn requires evaluation of the many-body electron wave function of the states $A$ and $B$, as in Ref.~\refcite{Hohensee2013}.  Here, we estimate the size of this matrix element by treating these states as hydrogenic wave functions of the leading order configurations for state $A$ ([Xe]$4f^{10}5d6s$), and state $B$ ([Xe]$4f^{9}5d^{2}6s$).  We assume that the relevant matrix element is that between the $4f$ and $5d$ orbitals, and find that $\langle 4f|c\partial_{x_{3}}|5d\rangle \sim 0.01\,ca_0^{-1} \approx 6\times10^{16}\,\mathrm{Hz}$, where $a_{0}$ is the Bohr radius. 

The magnitude of the parity violating matrix element $H_{w}$ in Dy was constrained to be $|H_w| \lesssim 5$~Hz in Ref.~\refcite{Nguyen1997}, limited by the available statistics.  A search for orientation-dependent variations in the parity-violating matrix element with the same precision could constrain the $c_{(0j)}$ terms at the level of $10^{-16}$, which would be competitive with astrophysical bounds~\cite{datatables}. Note that this estimate neglects possible enhancements due to the large nuclear charge, $Z = 66$, of Dy.  Using the apparatus of Ref.~\refcite{Hohensee2013}, with minimal changes, a new experiment should shrink this statistical limit by a factor of $>50$.  This could allow us to detect parity-odd, $c_{(0j)}$ coefficients as small as $10^{-18}$ with a low-energy tabletop experiment.  We caution that this is contingent on our estimate of the matrix elements $\left\langle\left.\left. A\right|W_{j}\right|B\right\rangle$ for dysprosium.  A full calculation of these matrix elements will be the subject of future work.

We thank B. Altschul, D. Budker, V. Flambaum, A. Kosteleck\'{y}, H. M\"uller, and J. Tasson for stimulating discussions. N.L. would like to thank L. Hunter and D. Phillips for their positive reception and encouragement of this idea.  N.L. acknowledges the support of NSF grant PHY-1068875.

\end{document}